\documentstyle[11pt,epsf]{article}
\makeatletter
\newcommand{\artsectnumbering}{%
\@addtoreset{equation}{section}
\renewcommand{\theequation}{\thesection.\arabic{equation}}}
\makeatother  
\newcommand{\al}{\alpha}

\newcommand{\bv}{\begin{verbatim}}
\newcommand{\de}{\delta}
\newcommand{\ch}{\chi}
\newcommand{\De}{\Delta}

\newcommand{\ep}{\epsilon}

\newcommand{\ev}{\end{verbatim}}
\newcommand{\fr}{\frac}
\newcommand{\Ga}{\Gamma}
\newcommand{\ga}{\gamma}

\newcommand{\La}{\Lambda}
\newcommand{\la}{\lambda}

\newcommand{\om}{\omega}
\newcommand{\ph}{\phi}

\newcommand{\si}{\sigma}
\newcommand{\ta}{\tau}

\newcommand{\Th}{\Theta}
\newcommand{\th}{\theta}

\newcommand{\be}{\begin{equation}}
\newcommand{\ee}{\end{equation}} 
\newcommand{\eei}{\end{equation}\indent\indent}
\newcommand{\bc}{\begin{center}}
\newcommand{\ec}{\end{center}}
\newcommand{\ber}{\begin{eqnarray}}
\newcommand{\ear}{\end{eqnarray}}
\newcommand{\ba}{\begin{array}}
\newcommand{\ea}{\end{array}}

\newcommand{\p}{\partial}

\newcommand{\2}{^{(2)}}

\newcommand{\Lg}{{\cal L}}

\def\case#1/#2{\textstyle\frac{#1}{#2} }
\begin{document}
\title{The World Function in Robertson-Walker Spacetime.}
\author{Mark D. Roberts, \\\\
Department of Mathematics and Applied Mathematics, \\ 
University of Cape Town,\\
Rondbosch 7701,\\
South Africa\\\\
roberts@gmunu.mth.uct.ac.za} 
\date{\today}
\maketitle
\vspace{0.1truein}
\bc Published:  {\it Astro.Lett. and Communications} {\bf 28}(1993)349-357.\ec
\bc Eprint: gr-qc/9905005\ec
\bc Comments:  12 pages, no diagrams or tables,  LaTex2e.\ec
\bc 2 KEYWORDS:\ec
\bc World Function:~~~Robertson-Walker Spacetime.\ec
\bc 1999 PACS Classification Scheme:\ec
\bc http://publish.aps.org/eprint/gateway/pacslist \ec
\bc 04.20.Cv, 95.10-a\ec
\bc 1991 Mathematics Subject Classification:\ec
\bc http://www.ams.org/msc \ec
\bc 83F05,  83C10\ec
\newpage
\artsectnumbering
\begin{abstract}
A method for finding the world function of Robertson-Walker spacetimes is 
presented.   It is applied to find the world function for the $k=0,  \ga=2$,
solution.   The close point approximation for the Robertson-Walker world
function is calculated upto fourth order.
\end{abstract}
\section{Introduction}
\label{sec:intro}
A fundamental invariant of a spacetime is the world function,  
which is given by
\be
\si=\pm\fr{1}{2}\ta^2,
\label{eq:1.1}
\ee
where $\ta$ is the geodesic distance.   $\si$ is positive for timelike 
geodesics,  zero for null geodesics,  and negative for spacelike geodesics.
This invariant contains a lot of information about an arbitrary Riemannian 
manifold.   Once it is known the metric can be calculated from it using the 
coincidence limit of one point on the geodesic approaching the other,  thus
$\lim_{x'\rightarrow x}\ta_{;ab}=g_{ab}$,  and then the Riemann tensor can
be calculated from the metric.   Defining the Van Vleck determinant
$\De\equiv g^{1/2}(x)det(\si_{ab})g^{1/2}(x')$ a Hadmard Green function is
$G(x,x')=i(8\pi^2)^{-1}\De^{1\2}(\si+i\ep)^{-1}$,  thus the world function,
or an approximation to it,  is needed when this Green function is required.
These Green functions are used a lot in the theory of quantum fields in
curved spaces,  see for example Brown and Ottewill (1983) \cite{bi:brown};
and also they are used in the study of the motion of charged particles 
in curved spacetimes,  Roberts (1989) \cite{bi:mdr1}.

There are very few metrics for which the exact world function is known.
It has been found for:  DeSitter spacetime by Ruse (1930) \cite{bi:ruse},
for $pp$ waves by G\"unther (1965) \cite{bi:gunther},  for G\"odel's solution 
by Buchdahl and Warner (1980) \cite{bi:BW}, and for the Bertotti-Robinson 
solution by John (1984,1989) \cite{bi:john84} \cite{bi:john89}.   Buchdahl
(1972) \cite{bi:buchdahl} has studies the production of the world function in
Robertson-Walker spacetimes.   In section \ref{sec:3} one new exact world 
function is found for the $k=0,\ga=2$ Robertson-walker solution,  and 
previously known exact world functions for Robertson-Walker solutions 
collected together.   For the majority of Robertson-Walker metrics it is
impossible to find the world function because intractable problems with 
elliptical integrals occur.   For many applications it is only necessary to 
know the world function in the close point approximation and this is given in 
section \ref{sec:4}.   The conventions used are:  signature $+---$;  
$a,b,c=0,1,2,3$;  $i,j,k=1,2,3$;
\ber
R^a_{.bcd}&=&\p_c\Ga^a_{db}-\p_d\Ga^a_{cb}
              +\Ga^a_{cf}\Ga^f_{db}-\ga^a_{df}\Ga^f_{cb}\\
R_{bd}&=&R^a_{.bda}.\nonumber
\label{eq:1.2}
\ear

The Robertson-Walker line element is
\be
ds^2=N(t)^2dt^2-R(t)^2(d\ch^2+f(\ch)^2(d\th^2+sin^2(\th)d\ph^2))
\label{eq:1.3}
\ee
where $R$ is the scale factor and
\begin{equation}
f(\ch)=\left\{
\begin{array}{rcl}
sin(\ch)~~~&{\rm if}&~k=+1\nonumber\\
\ch~~~&{\rm if}&~k=0\\
sinh(\ch)~~~&{\rm if}&~k=-1\nonumber
\end{array}
\right.
\label{eq:1.4}
\end{equation}
in the conventions for $k$ of Hawking and Ellis (1973) \cite{bi:HE}p.136.

The lapse function $N$ is arbitrary,  and it can be absorbed into the line
element;  two common choices for $N$ are $N=1$,  and $N=R$ in which form the 
line element is conformal to the Einstein static universe.   
The Ricci tensor is given by
\ber
R_{tt}&=&\fr{3R_{,tt}}{R}-\fr{3N_{,t}R_{,t}}{NR},\nonumber\\
R_{ij}&=&\fr{g_{ij}}{R^2}\left(-2k-\fr{RR_{,tt}}{N^2}-\fr{2R^2_{,t}}{N^2}
                               +\fr{RR_{,t}N_{,t}}{N^3}\right).
\label{eq:1.5}
\ear
Using Einstein's field equations the density $\mu$ and the pressure $p$ for a
perfect fluid are given by
\ber
&&8\pi G\mu=\fr{3k}{R^2}+\fr{3R^2_{,t}}{N^2R^2},\nonumber\\
&&8\pi Gp  =-\fr{k}{R^2}-\fr{R^2_{,t}}{N^2R^2}+\fr{2N_{,t}R_{,t}}{N^3R}
                              +\fr{2R_{,tt}}{N^2R},
\label{eq:1.6}
\ear
solutions to these equations with the equation of state
\be
p=(\ga-1)\mu,
\label{eq:1.7}
\ee
have been given by Vajk (1969) \cite{bi:vajk}.   The speed of sound in a
fluid is given by $c~dp/d\mu$,  Hawking and Ellis (1973) \cite{bi:HE}p.91,
thus the speed of sound for the equation of state \ref{eq:1.7} is
\be
c(\ga-1).
\label{eq:1.8}
\ee
\section{The Geodesics of Robertson-Walker Spacetime}
\label{sec:2}
The geodesic equations can be most elegantly derived by the Euler method,
the Lagrangian is
\be
2{\mathcal L}=N^2\dot{t}^2-R^2g_{ij}^{(3)}\dot{x}^i\dot{x}^j,
\label{eq:2.1}
\ee
where:(i)
\be
\dot{x}^a=\fr{dx^a}{d\ta}
\label{eq:2.2}
\ee
(ii) $\ta$ is the geodesic distance,\\
(iii) $2{\mathcal L}=1,0, or -1$ for timelike,  null,  
or spacelike geodescs respectively,\\
(iv) $g_{ij}^{(3)}$ is the metric of the three space of constant curvature,\\
and $i,j=1,2,3$.   The momenta are
\ber
&&p^t=\fr{\p {\mathcal L}}{\p \dot{t}}=N^2\dot{t}\\
&&p_i=-\fr{\p {\mathcal L}}{\p \dot{x}^i}=R^2\dot{x}_i,\nonumber
\label{eq:2.3}
\ear
which give
\ber
&&\fr{dp_t}{d\ta}=2NN_{,t}\dot{t}^2+N^2\ddot{t}\nonumber\\
&&\fr{dp_i}{d\ta}=2RR_{,t}\dot{t}\dot{x}_i+R^2\ddot{x}_i.
\label{eq:2.4}
\ear
The Lagrangian gives
\ber
&&\fr{\p {\mathcal L}}{\p t}=NN_{,t}\dot{t}^2
                             -RR_{,t}g_{ij}^{(3)}\dot{x}^i\dot{x}^j\nonumber\\
&&\fr{\p {\mathcal L}}{\p x^i}=-R^2g_{kj,i}^{(3)}\dot{x}^k\dot{x}^{j}\dot{x}^i.
\label{eq:2.5}
\ear
The Euler equation
\be
\fr{\p {\mathcal L}}{\p x^i}=\fr{d p^t}{d\ta},
\label{eq:2.6}
\ee
after dividing by $N^2$ becomes
\be
\ddot{t}+\fr{N_{,t}}{N}\dot{t}^2
+\fr{RR_{,t}}{R^2}g_{ij}^{(3)}\dot{x}^i\dot{x}^j=0.
\label{eq:2.7}
\ee
The Euler equation
\be
-\fr{\p {\mathcal L}}{\p x^i}=\fr{d p_i}{d\ta},
\label{eq:2.8}
\ee
after dividing by $R^2$ becomes
\be
\ddot{x}_i+\fr{2R_{,t}}{R}\dot{t}\dot{x}_i
   -g_{kj,i}\dot{x}^k\dot{x}^j\dot{x}^i=0.
\label{eq:2.9}
\ee

The geodesic distance $\om$ on the three space of constant curvature
is given by the well known expression
\ber
cos(\om)&=&cos(\ch)cos(\ch')
          +sin(\ch)sin(\ch')cos(\Th)~~~~~~~~~{\tt if}~ k=+1\nonumber\\
\om^2&=&\ch^2+\ch'^2
          -2\ch \ch'cos(\Th)~~~~~~~~~~~~~~~~~~~~~~~~~~~{\tt if}~ k=0\\
cosh(\om)&=&cosh(\ch)cosh(\ch')
           +sinh(\ch)sinh(\ch')cos(\Th)~~~{\tt if}~ k=-1,\nonumber
\label{eq:2.10}
\ear
where
\be
cos(\Th)=cos(\th)cos(\th')+sin(\th)sin(\th')cos(\ph-\ph').
\label{eq:2.11}
\ee
Now
\be
g_{ij}^{(3)}\dot{x}^i\dot{x}^j=\left(\fr{d\om}{d\ta}\right)^2
   \fr{dx^i}{d\om}\fr{dx^j}{d\om}=\dot{\om}^2
\label{eq:2.12}
\ee
Equations \ref{eq:2.1} and \ref{eq:2.12} give
\be
2{\mathcal L}=N^2\dot{t}^2-R^2\dot{\om}^2,
\label{eq:2.13}
\ee
and Equations \ref{eq:2.7} and \ref{eq:2.8} give
\be
\ddot{t}+\fr{N_{,t}}{N}\dot{t}^2+\fr{RR_{,t}}{N^2}\dot{\om}^2=0.
\label{eq:2.14}
\ee
Noting that
\be
(g_{kj}^{(3)}\dot{x}^k\dot{x}^j)_{,i}=(\dot{\om})_{,i}=0,
\label{eq:2.15}
\ee
Equation \ref{eq:2.9} becomes
\be
\ddot{\om}+2RR_{,t}\dot{t}\dot{\om}=0.
\label{eq:2.16}
\ee
The geodesics are determined by \ref{eq:2.13},  \ref{eq:2.14},  
and \ref{eq:2.16};   \ref{eq:2.14} can be obtained by differentiating 
\ref{eq:2.13} and substituting \ref{eq:2.16},  therefore only equations
\ref{eq:2.13} and \ref{eq:2.16} are used.   Equation \ref{eq:2.16} can be 
integrated to give
\be
\dot{\om}=\fr{\al}{R^2}
\label{eq:2.17}
\ee
where $\al$ is a constant independent of $\ta$.
Rearranging \ref{eq:2.13} and substituting \ref{eq:2.17} gives
\be
N\dot{t}=\pm\sqrt{R^2\dot{\om}^2+2\Lg}
        =\pm\sqrt{\fr{\al^2}{R^2}+2\Lg}
\label{eq:2.18}
\ee
thus the dynamics of the geodesics are determined by \ref{eq:2.17} 
and \ref{eq:2.18}.   the integeral form of these equations are
\ber
\ta&=&\pm\int\fr{RN~dt}{\sqrt{\al^2+2\lg R^2}},\label{eq:2.19}\\
\om&=&\pm\int\fr{\al N~dt}{R\sqrt{\al^2+2\Lg R^2}},\label{eq:2.20}
\ear
takeing $2\Lg=+1$ for timelike geodesics,  these are just the equations 
describing geodesics given by Robertson (1933) \cite{bi:robertson}.
$\ta$ is evaluated from \ref{eq:2.19} with the positive sign and the limits
of integration from $T'$ to $t$ and with $\al$ replaced by its value
evaluated from \ref{eq:2.20}.  The the world function $\si$ is given from $\ta$
by \ref{eq:1.1}.
\section{Examples of exact World Functions for Robertson-Walker Spacetime.}
\label{sec:3}
{\it Example} 1  The Einstein Static Universe.
The metric is given by \ref{eq:1.3} with
\be
N=1,~~~R=R_0,
\label{eq:3.1}
\ee
where $R_0$ is a constant.   The ricci tensor is given by
\ber
R_{tt}&=&0\nonumber\\
R_{ij}&=&-\fr{2kg_{ij}}{R_0^2},
\label{eq:3.2}
\ear
thus for $f=0$ the spacetime is flat.
The pressure and the density are given by
\be
8\pi G\mu=-24\pi Gp=\fr{3k}{R_0^2},
\label{eq:3.3}
\ee
therefore the equation of state is \ref{eq:1.7} with $\ga=\fr{2}{3}$.
the world function is given by
\be
2\si=(t-t')-R_0^2\om^2
\label{eq:3.4}
\ee
{\it Example} 2 The Generalized Milne Universe.   For small time variation
the metric can be expanded as a Taylor series in $t$
\ber
N&=&1\nonumber\\
R&=&R_0[1+H_0t-\fr{1}{2}q_0H_0^2t^2=O(t^3)]
\label{eq:3.5}
\ear
where $H_0$ is the Hubble constant and $q_0$ is the deceleration parameter.
The $q_0\ne0$ case proves to be intractable,  so that only terms up to first
order are considered
\ber
N&=&1\nonumber\\
R&=&R_0[1+H_0t].
\label{eq:3.6}
\ear
The Ricci tensor is
\ber
&&R_{tt}=0\nonumber\\
&&R_{ij}=-2g_{ij}\fr{(R_0^2H_0^2+k)}{R_0^2(!=H_0t)^2},
\label{eq:3.7}
\ear
thus when $k=-1$ and $R_0H_0=1$ the spacetime is flat,
and called the Milne Universe,
which is just Minkowski spacetime with a point removed at the origin.
The pressure and density are given by
\be
8\pi G\mu=-24\pi Gp=\fr{3(k+R_0^2H_)^2)}{R_0^2(1+H_0t)^2}
\label{eq:3.8}
\ee
therefore the equation of state is \ref{eq:1.5} with $\ga=\fr{2}{3}$.
Gefining the new time coordinate
\be
\bar{t}=\fr{1}{H_0}+t,
\label{eq:3.9}
\ee
the metric is given by
\ber
N&=&1,\nonumber\\
R&=&R_0H_0\bar{t}.
\label{eq:3.10}
\ear
The world function is given by
\be
2\si=\bar{t}^2+\bar{t}'^2-2\bar{t}\bar{t}'cosh(R_0H_0\om).
\ee
Transforming to the original coordinate system \ref{eq:3.6} this is
\be
2\si=(t+t')^2-4\left(\fr{1}{H_0}
   +t\right)\left(\fr{1}{H_0}+t'\right)sinh^2(\fr{1}{2}R_0H_0\om)
\label{eq:3.12}
\ee

{\it Example 3} DeSitter and Anti DeSitter Spacetimes.
\footnote {footnote added 1999,  a more complete treatment of the world 
function in DeSitter and Anti DeSitter spacetimes can be found in my 
work ``Two Point Gravitational Energy'' in preparation.}
These are maximally symmetric solutions to
\be
R_{ab}=-\La g_{ab}
\label{eq:3.13}
\ee
with $\La>0$ for DeSitter spacetime and $\La<0$ for Anti DeSitter spacetime.
For DeSitter spacetime the metric pressure and density are given by
\ber
k=0,~~~
N=R=\fr{1}{\la t},~~~
\la=\sqrt{\fr{\La}{3}},\\
8\pi G\mu=-8\pi Gp=\La,\nonumber
\label{eq:3.14}
\ear
and therefore the equation of state is given by \ref{eq:1.7} with $\ga=0$.
For DeSitter spacetime the world funcion has been found by Ruse (1930)
\cite{bi:ruse} and is
\be
\ta=\sqrt{2\si}=\fr{1}{\la}~ arc~ cos(Q),
\label{eq:3.15}
\ee
where
\be
Q=\fr{1}{2tt'}(t^2+t'^2\om^2)
\label{eq:3.16}
\ee
is the geometric distance in the five dimensional embedding space,
and obeys the differential equation
\be
Q_aQ^a=\la^2(1-Q^2).
\label{eq:3.17}
\ee
Transferring coordinates the result for DeSitter spacetime can be expressed as
\ber
k=+1,~~~
N=1,~~~
R=\fr{1}{\la}cosh(\la t),~~~
\la=\sqrt{\fr{-\La}{3}},\\
cosh(\ta)=sinh(\la t)sinh(\la t')+cosh(\la t)cosh(\la t')cos(\om/\la),\nonumber
\label{eq:3.18}
\ear
by symmetry the world function for Anti DeSitter spacetime can be found to be
\ber
k=-1,~~~
N=1,~~~
R=\fr{1}{\la}cos(\la t),~~~
\la=\sqrt{\fr{\La}{3}},\\
cos(\ta)=sin(\la t)sin(\la t')+cos(\la t)cos(\la t') cosh(\om/\la).\nonumber
\label{eq:3.19}
\ear

{\it Example 4}.  Robertson-Walker Spacetime with a Massless Scalar Field 
as Source.   The world function for this spacetime is new and is derived
here in detail,  thus illustrating the method of Section \ref{sec:2}.
The metric is given by \ref{eq:1.3} with
\be
N=R=bt^{1/2},
\label{eq:3.20}
\ee
where $b$ is a constant.   The Ricci tensor is given by
\ber
R_{tt}&=&\fr{-3}{2t^2},\nonumber\\
R_{ij}&=&\fr{-2}{b^2t}g_{ij}.
\label{eq:3.21}
\ear
The pressure and density are
\ber
8\pi G\mu&=&\fr{3}{b^2t}\left(k+\fr{1}{4t^2}\right),\nonumber\\
8\pi Gp  &=&\fr{1}{b^2t}\left(k+\fr{3}{4t^2}\right),
\label{eq:3.22}
\ear
thus when $k=o$ the equation of state is just \ref{eq:1.7} with $\ga=2$,
for which the speed of sound in the fluid equals 
the speed of light in a vacuum.

Such a fluid is equivalent to a massless scalar field,
see for example Roberts (1989) \cite{bi:mdr2},  with a Ricci tensor
\be
R_{ab}=-2\ph_a\ph_b,
\label{eq:3.23}
\ee
for \ref{eq:3.21} $\ph$ is given by
\be
\ph=\fr{1}{2}\sqrt{3}ln(t).
\label{eq:3.24}
\ee
The integrals \ref{eq:2.19} and \ref{eq:2.20} become
\ber
\ta&=&\int^t_{t'}\fr{b^2t~dt}{\sqrt{\al^2+b^2 t}}
   =\fr{2}{3}\sqrt{\al^2+b^2 t}\left(t-\fr{2\al^2}{b}\right)|^t_{t'},
\label{eq:3.25}\\
\om&=&\int^t_{t'}\fr{\al~dt}{\sqrt{\al^2+b^2t}}
   =\fr{2\al}{b^2}\sqrt{\al^2+b^2t}|^t_{t'}
\label{eq:3.26}
\ear
Squaring \ref{eq:3.26} gives
\be
\fr{b^4\om^2}{4\al^2}-2\al^2-b^2(t+t')=-2\sqrt{\al^2+b^2t}\sqrt{\al^2+b^2t'},
\label{eq:3.27}
\ee
Squaring \ref{eq:3.27} and multiplying by $\al^4b^{-4}$ 
gives the quadratic in $\al^2$
\be
\fr{\om^4b^4}{16}-\fr{\al^2\om^2b^2}{4}(t+t')+\al^4((t-t')^2-\om^2)=0,
\label{eq:3.28}
\ee
which has solution
\be
\al^2=\fr{1}{8}\si_e^{-1}b^2\om^2(t+t'\pm\sqrt{4tt'+\om^2}),
\label{eq:3.29}
\ee
where
\be
2\si_e=(t-t')^2-\om^2
\label{eq:3.30}
\ee
then from \ref{eq:3.25} the world function is given by
\ber
\fr{3}{2b}\sqrt{2\si}
=\fr{3}{2b}\ta
&=&(\fr{1}{8}\si_e^{-1}\om^2(t+t'\pm\sqrt{4tt'+\om^2})+t)^{1/2}\nonumber\\
&&\times(t-\fr{1}{4}\si_e^{-1}\om^2(t+t'\pm\sqrt{4tt'+\om^2}))\nonumber\\
&&-(\fr{1}{8}\si_e^{-1}\om^2(t+t'\pm\sqrt{4tt'+\om^2}+t')^{1/2}\nonumber\\
&&\times(t'-\fr{1}{4}\si_e^{-1}\om^2(t+t'\pm\sqrt{4tt'+\om^2}))
\label{eq:3.31}
\ear

{\it Example} 5 The Tolman Universe.   For the Tolman radiation Universe the
metric is given by \ref{eq:1.3} with
\ber
&&R=N=bt,\\
&&k=o,\nonumber\\
&&\ga=\fr{4}{3}.\nonumber
\label{eq:3.32}
\ear
The integrals \ref{eq:2.19} and \ref{eq:2.20} can be evaluated to give
\be
\ta=\left[\fr{\al t}{2}\sqrt{1-\fr{b^2t}{\al^2}}
       -\fr{\al^2}{2b}~arc~sinh(\fr{bt}{\al}\right]^t_{t'}
\label{eq:3.33}
\ee
\be
\om=\fr{\al}{b}~arc~sinh(\fr{bt}{\al})|^t_{t'},
\label{eq:3.34}
\ee
and \ref{eq:3.34} cannot be inverted to give $\al$.   
For the majority of solutions with equation of state \ref{eq:1.7},
the integrals \ref{eq:2.19} and \ref{eq:2.20} are elliptical and the problem
becomes intractable.   The Tolman universe illustrates that the occurrence 
of elliptical integrals is not the only problem that might arise;
because in this case the integrals can be evaluated to give simple functions
for $\ta$ and $\om$,  however an expression for $\al$ cannot be extracted
from these functions.
\section{The Close Point Approximation}
\label{sec:4}
Introducing the notation
\be
\pi^{ab\ldots c}=(x^a-x'^a)(x^b-x'^b)\ldots(x^xc-x'^c),
\label{eq:41.}
\ee
and
\be
\pi^n=\pi^a_1\ldots \pi^a_n,
\label{eq:4.2}
\ee
the world function can be expressed in the form
\be
2\si=g_{ab}\pi^{ab}+h_{abc}\pi^{abc}+O(\pi^4).
\label{eq:4.3}
\ee
Differentiating
\be
2\si_e=2g_{ae}\pi^a+\pi^{ab}(g_{ab,e}+h_{abe}+h_{aeb}+h_{eab})+O(\pi^3),
\label{eq:4.4}
\ee
thus
\be 
\si_e\si^e=g_{ab}\pi^{ab}+\fr{1}{2}(g_{ab,c}+3h_{abc})\pi^{abc}+O(\pi^4),
\label{eq:4.5}
\ee
now because
\be
2\si=\si_e\si^e,
\label{eq:4.6}
\ee
Equations \ref{eq:4.3} and \ref{eq:4.5} can be equated to give
\be
h_{abc}=-\fr{1}{2}g_{ab,c}.
\label{eq:4.7}
\ee
Carrying out the above approximation scheme to fourt order gives
\ber
2\si&=&g_{ab}\pi^{ab}-\fr{1}{2}g_{ab,c}\pi^{abc}\nonumber\\
+&\fr{1}{12}&[2g_{ab,cd}-g^{ef}(\fr{1}{2}g_{ab,e}-g_{ae,b})
      (\fr{1}{2}g_{cd,f}-g_{cf,d})]\pi^{abcd}+O(\pi^5).
\label{eq:4.8}
\ear
For the line element \ref{eq:1.3} this gives
\ber
2\si&=&N^2\de t^2-R^2\om^2-NN_{,t}\de t^3+RR_{,t}\de t \om^2\\
&&\fr{1}{12}(3N_{,t'}^2+4NN_{,tt})\de t^4
  -\fr{1}{3}\left(RR_{,tt}+\fr{RN_{,t}R_{,t}}{N}\right)\de t^2\om^2\nonumber\\
&&-\fr{1}{12}\left(\fr{RR_{,t}}{N}\right)^2\om^4+O(\pi^5),\nonumber
\label{eq:4.9}
\ear
where
\be
\de t=t-t',
\label{eq:4.10}
\ee
and $\om$ is given by \ref{eq:2.10}.
\section{Acknowledgement}
I would like to thank the Leverhulme Trust for financial support.

\end{document}